\documentclass[12pt]{article}
\usepackage{graphicx}
\usepackage{amsmath}
\usepackage{amsfonts}
\usepackage{amssymb}
\begin{document}
\input psfig.sty

\begin{center}
GIANT PEAK OF THE 1/f NOISE AT THE METAL-INSULATOR 
TRANSITIONS IN LOW-$T_{c}$ CMR MANGANITES: \\
EVIDENCE OF THE PERCOLATION THRESHOLD AT $T_{c}$
\end{center}

\vspace{0.25 truein} \noindent V. PODZOROV$^{1}$, M. UEHARA$^{1}$, M. E.
GERSHENSON$^{1}$, and S-W. CHEONG$^{1,2}$

\noindent1) Serin Physics Laboratory, Rutgers University, Piscataway, NJ 08854-8019

\noindent2) Bell Labs, Lucent Technologies, Murray Hill, NJ 07974

\vspace{0.375 truein}

\noindent ABSTRACT \vspace{0.25 truein}

We observed a dramatic peak in the $1/f$ noise at the metal-insulator
transition (MIT) in low-$T_{c}$ manganites. This many-orders-of-magnitude
noise enhancement is observed for both polycrystalline and single-crystal
samples of $La_{5/8-y}\Pr_{y}Ca_{3/8}MnO_{3}$ ($y=0.35-0.4$) in zero magnetic 
field and $\Pr_{1-x}Ca_{x}MnO_{3}$ ($x=0.35-0.5$) in magnetic field. 
This observation strongly suggests that the microscopic
phase separation in the low-$T_{c}$\ manganites causes formation of a
percolation network, and that the observed MIT is a percolation
threshold.
\vspace{0.25in}

\noindent INTRODUCTION \vspace{0.25in}

It is well known that the electronic phase diagrams of perovskite manganites
are very complex; they exhibit numerous ground states and phase transitions as
the carrier concentration is varied \cite{imada}. The phase diagram for the
system $La_{1-x}Ca_{x}MnO_{3}$ in the plane of the doping concentration
$x$\ and temperature $T$ is reported in \cite{cheonghwang}. The concentration of \ charge
carriers in this system is proportional to the $Ca$ doping level. At high
temperatures, the system is in a paramagnetic phase. In the range $x\simeq0.3-0.5$, the
system undergoes a transition into the ferromagnetically-ordered state at
$T\simeq200-250K$; this magnetic transition is accompanied by the
metal-insulator transition (MIT) \cite{helmholt,jin,nagaev}. A very high
sensitivity of this MIT to the external magnetic field results in the
so-called colossal magnetoresistance (CMR) \cite{helmholt,jin}. The change of
the resistivity across the transition and, correspondingly, the CMR are more
dramatic in compounds with a lower transition temperature $T_{c}$. The MIT in
low-$T_{c}$ manganites bears many features which are intrinsic to the
first-order transitions, including a strong thermal hysteresis of the
resistivity $\rho$ and magnetization $M$ \cite{babushkina,masatomo}.

There is a growing theoretical and experimental evidence that transport
properties of the insulating state above $T_{c}$ are dominated by small
polarons or magnetic polarons, and that the band-like carriers become
important below $T_{c}$ \cite{millis1,roder,zhou}. However, the details of the
MIT remain unclear. The experimental data suggest that the phase separation
occurs \textit{gradually} with decreasing temperature \cite{babushkina}. 
Appearance of small ferromagnetic (FM) regions
in the charge-ordered (CO) phase has been reported at $T>>T_{c}$, well beyond
the conventional fluctuation regime \cite{De Teresa}.\ A very large and
\textit{temperature-independent} resistivity in the FM state, much greater
than the Ioffe-Regel limit for a uniform metallic system, also indicates that
the insulating CO and metallic FM phases coexist on the ''metallic'' side of
the MIT \cite{babushkina,masatomo}. The transition into the FM state is
accompanied by an increase of the magnetization $M$, which saturates below
$T_{c}$ (see Fig. 1). In contrast to the resistance, $M$ varies
smoothly across the transition. The magnetization is proportional to the
volume fraction of the FM phase; this volume fraction is $\sim15-20\%$ at
$T_{c}$, which is close to the percolation threshold in a three-dimensional
percolated system. All these observations suggest that the low-$T_{c}$ manganites can be
viewed as macroscopically inhomogeneous systems, where the metallic FM domains are
imbedded into the insulating CO matrix. Percolation phenomena might be
important in this situation near the MIT, provided the scale of the phase
separation is much smaller than the sample's dimensions.

\begin{figure}[tb]
\begin{center}
\includegraphics[height=5in, width=4in]{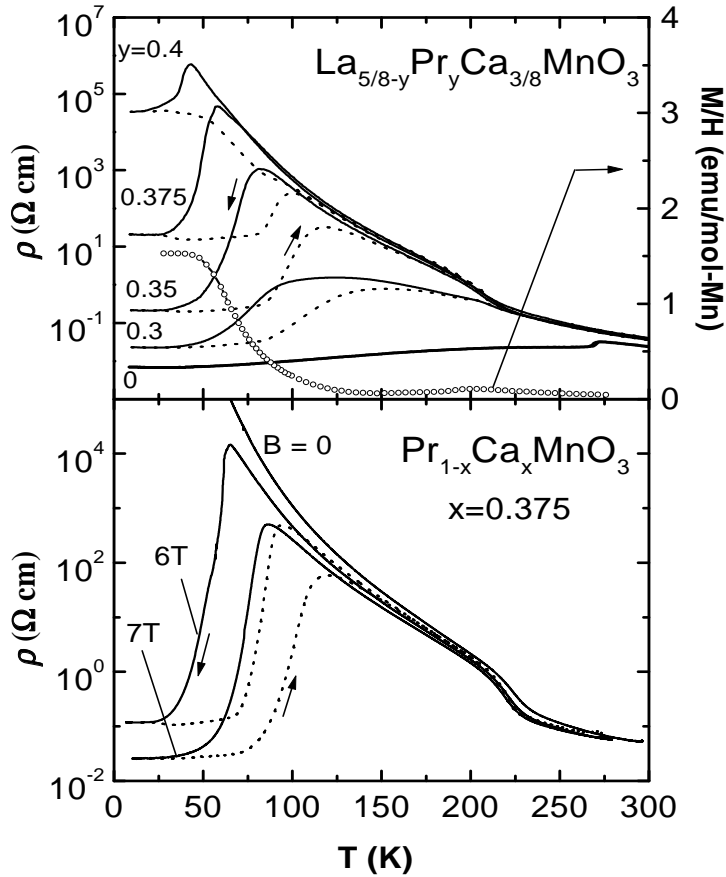}
\caption{Temperature dependence of the resistivity in $La_{5/8-y}%
Pr_{y}Ca_{3/8}MnO_{3}$ (at $B=0$) and in $\Pr_{1-x}Ca_{x}MnO_{3}$ (at
$B=0-7T$). The solid lines correspond to cooling, the dashed ones - to
heating. The temperature dependence of the magnetization for $La_{5/8-y}%
Pr_{y}Ca_{3/8}MnO_{3}$ with $y=0.35$ is shown for cooling as open circles in
the upper panel.} 
\label{Fig.1}
\end{center}
\end{figure}

The noise measurements can open a new window on the MIT in the CMR manganites.
Indeed, it is well known that in classical percolation systems, the $1/f$
noise diverges at the percolation threshold \cite{rammal1,kogan}. The $1/f$ noise
reflects fluctuations of the resistance, and its spectral density $S_{V}$ is
proportional to the fourth power of the bias current density $j$ \cite{kogan}. Close to the
percolation threshold, the current density becomes strongly non-uniform, and,
with approaching the threshold, the contribution of the regions with a large
$j$ to the $1/f$ noise increases more rapidly than their contribution to the
resistance \cite{kogan}. We report on the $1/f$ noise measurements in
polycrystalline and single crystal samples of low-$T_{c}$ manganites. The
combined transport and noise measurements strongly suggest that the so-called
Curie temperature in the low-$T_{c}$ materials is, in fact, a percolation
threshold temperature rather than the temperature of the long-range
ferromagnetic phase transition. The scaling analysis of the $1/f$ noise is
consistent with the percolation model of conducting domains randomly
distributed in an insulating matrix \cite{tremblay}.

\vspace{0.25in} \noindent EXPERIMENT \vspace{0.25in}

The resistance drop at the MIT is more dramatic in the CMR manganites with a
lower transition temperature $T_{c}$. One can \ decrease $T_{c}$, for example,
by partial substitution of $La$ with $\Pr$, which changes the chemical
pressure in the system \cite{cheonghwang}. Alternately, it is possible to
start with a compound that does not demonstrate the MIT even at the lowest $T$
(e.g., $Pr_{1-x}Ca_{x}MnO_{3}$), and to induce the MIT by the magnetic field.
In this work, we have used both methods. The transport and noise measurements
have been carried out on poly- and single crystal bulk samples of
$La_{5/8-y}Pr_{y}Ca_{3/8}MnO_{3}$ ($y=0.35-0.4$) and $\Pr_{1-x}Ca_{x}MnO_{3}$
($x=0.25-0.5$). The former compound demonstrates the MIT at zero magnetic
field, whereas in the latter compound, the MIT can be induced by the magnetic
field (Fig. 1). The sample preparation is described
elsewhere \cite{cheonghwang}. Typically, the polycrystalline samples were
$4\times1\times1$ $mm^{3}$, single crystals - $3\times1\times0.5$ $mm^{3}$.
The spectral density of the $1/f$ noise and the resistivity $\rho$ have been
measured in the four-probe configuration over the temperature range
$T=4.2-300K$ in the magnetic field $B=0-8T$ \ for both cooling and heating.

The bias current $I$ was driven through the sample by a low-noise current
source with the output resistance much greater than the sample's resistance.
Typical values of the bias current were $I\simeq10^{-6}-10^{-4}A$. All the
data discussed below were obtained in the linear regime, where the rms noise
was linear in current. The noise signal was amplified by a preamplifier PAR
113 and measured by a lock-in amplifier SR 830 in the mean average deviation
mode. This regime allows to measure continuously the spectral noise density
$S_{V}=\langle(V-\bar{V})^{2}$ $\rangle$ ( $\bar{V}$ is the average value of
the voltage across the sample, $\langle..\rangle$ stands for the time
averaging), while the temperature (or the magnetic field) is slowly varying.
Because of \ a very strong temperature dependence of the resistance in
manganites, especially in the vicinity of the MIT, the rate of the $T$ (or
$B$) sweep must be very slow (otherwise, the change of \ $\bar{V}$ during the
measurement time will contribute to $S_{V}$). This also requires a good
temperature stabilization (the long-term temperature stability in our
measuring set-up was better than $1mK$ at $T=10-300K$). Typically, the noise
was measured at $f=10-30Hz$, with the equivalent noise bandwidth $\Delta f=1-5Hz$
and the time constant $\tau=0.1-1s$. It has
been verified that the noise spectrum has a power-law form $1/f$\ $^{\gamma}$
in the frequency range $f=1-10^{3}Hz$ with $\gamma$\ close to unity for all temperatures.

\vspace{0.25in} \noindent RESULTS \vspace{0.25in}

Typical temperature dependences of the resistivity for the $LaPrCaMnO$ system
at $B=0$\ and for the $PrCaMnO$ system at different magnetic fields are shown
in Fig. 1. Qualitatively, the dependences $\rho (T)$ for these systems are
similar: with cooling, the resistivity grows exponentially below the CO
transition ($T_{CO}\sim200-220K$), decreases rapidly when the system undergoes
the MIT, and remains almost temperature-independent in the FM state. The
temperature of the MIT is strongly $y$-dependent: $T_{c}$ increases from $35K$
for $y=0.4$ to $75K$ for $y=0.35$. For the $LaPrCaMnO$ system, the change of
the resistivity is the most dramatic at $y=0.35-0.375$, the ''sharpness'' of
the transition is also the largest for these concentrations of $Pr$. Both
systems exhibit a very strong hysteresis upon cooling and heating. The
transition into the FM state is accompanied by the increase of the
magnetization $M$, which saturates at $T<T_{c}$ (see Fig. 1). In contrast to
$\rho$, $M$ changes gradually across $T_{c}$ for both poly- and single crystal
samples. According to the magnetization data, the fractional volume of the FM
phase, on the one hand, always exceeds $\sim20\%$ on the ''metallic'' side of
the MIT (we recall that the percolation threshold for both discrete and
continuum percolation models is $\sim17\%$ in three dimensions), on the other
hand, is significantly smaller than 100\%. This observation, as well as an
anomalously large and temperature-independent resistivity in the FM state
(observed even for single crystal samples), is consistent with the idea of
coexistence of the FM and CO phases on the ferromagnetic side of the MIT.

\begin{figure}[!thb]
\begin{center}
\includegraphics[height=5in, width=4in]{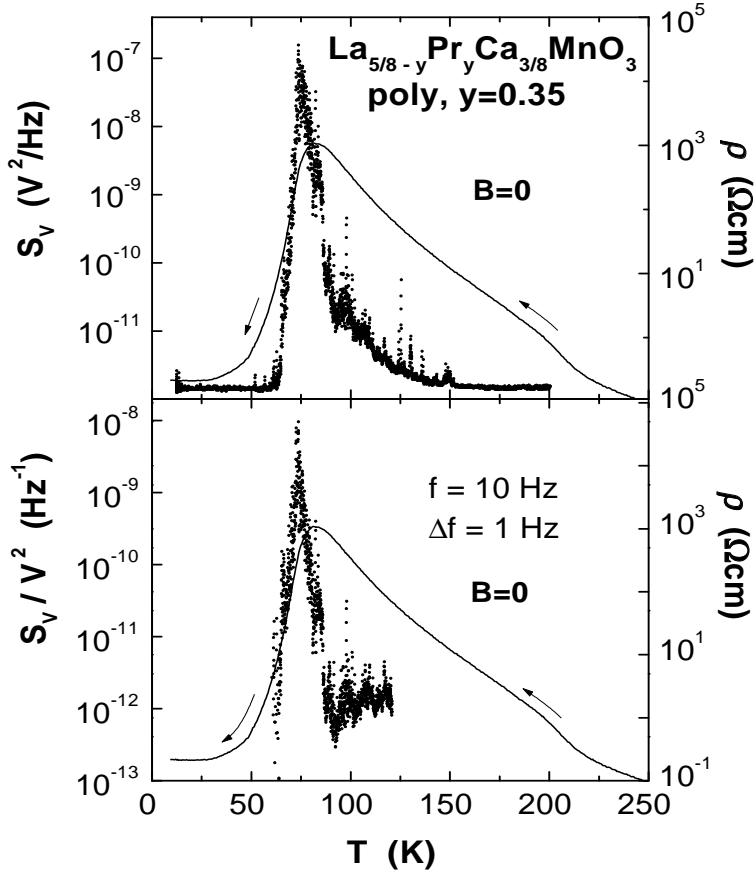}
\caption{Temperature dependences of the resistivity (solid lines), the spectral noise
density (dots, the upper panel), and the normalized spectral noise density corrected
on the background (dots, the lower panel) for a polycrystalline sample
$La_{5/8-y}Pr_{y}Ca_{3/8}MnO_{3}$ with $y=0.35$. All the dependences have been
measured for the zero-field cooling.} 
\label{Fig.2}
\end{center}
\end{figure}

The upper panel of Fig. 2 shows a typical temperature dependence of the
spectral noise density, measured for a polycrystalline sample $La_{5/8-y}%
Pr_{y}Ca_{3/8}MnO_{3}$ with $y=0.35$. The noise exceeds the background level
at $T\simeq150K$ (the preamplifier was not used in this measurement, which
explains a relatively large magnitude of the background), increases by \ many
orders of magnitude with approaching the MIT, and drops sharply below the
background level on the FM side of the transition. The frequency dependence of
this noise is close to $1/f$ . To compare the $1/f$ \ noise for different
samples, it is convenient to use the normalized spectral noise density,
$S_{V}/V^{2}=\langle(\delta\rho/\rho)^{2}\rangle$: in the linear regime, this
quantity does not depend on the bias current and on the sample's geometry. The
lower panel of Fig. 2 shows that $S_{V}/V^{2}$ for this sample is weakly
$T$-dependent in the CO state far from the MIT: it varies by a factor of 2-3
over the range $T=90-150K$, though $\rho$ changes by 2-3 orders of magnitude
over the same $T$ interval (see also the lower panel of Fig. 3). With
approaching $T_{c},$ the $1/f$ noise exhibits a giant peak. The peak value of
$S_{V}/V^{2}=10^{-10}-10^{-8}$ $Hz^{-1}$ exceeds by many orders of magnitude
$S_{V}/V^{2}$ observed for macroscopically uniform metallic or semiconducting
samples ($10^{-19}-10^{-23}$ $Hz^{-1}$ and $10^{-16}-10^{-19}$ $Hz^{-1}$,
correspondingly, for the same frequency and sample's volume) \cite{kogan}.

\begin{figure}[!hbt]
\begin{center}
\includegraphics[height=5in, width=4in]{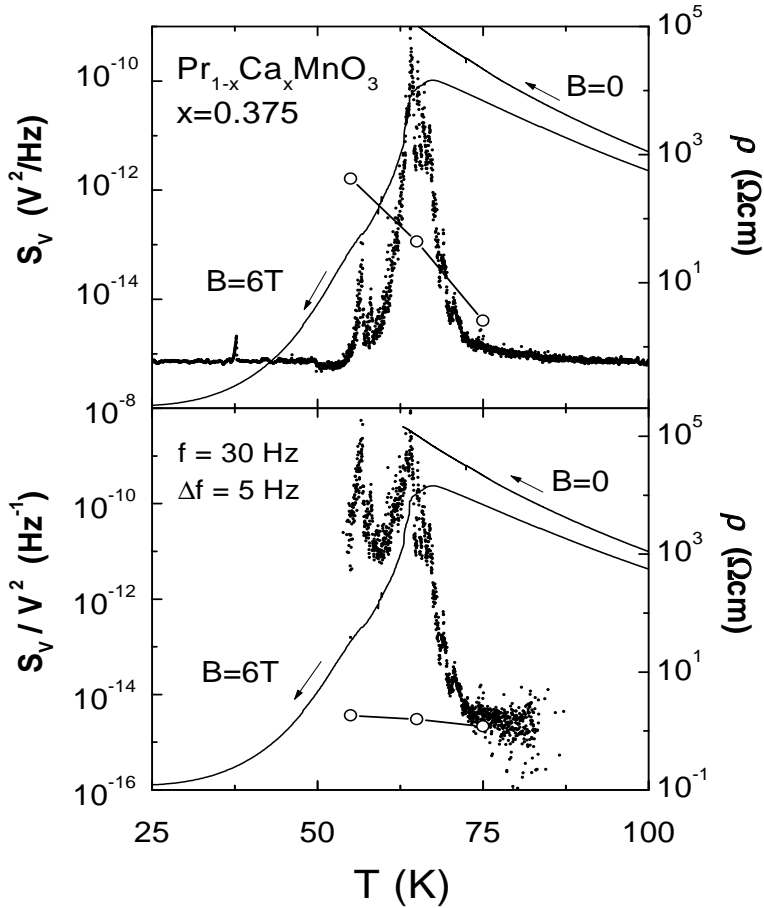}
\caption{Temperature dependences of the resistivity (solid lines), the spectral
noise density (dots and open circles, the upper panel), and the normalized spectral noise density
corrected on the background (dots and open circles, the lower panel) for a polycrystalline sample
$Pr_{1-x}Ca_{x}MnO_{3}$ with $x=0.375$. All the dependences have been measured
for cooling. At $B=0$, this compound does not exhibit the MIT. The normalized $1/f$ noise
measured for several fixed $T$ at $B=0$ coincides with the data
at $B=6T$, measured with a slow temperature sweep far from the transition. 
This indicates that the sweep rate ($4mK/s$) was
sufficiently low to exclude contribution of the time-dependent $\bar{V}$ \ to
$S_{V}=\langle(V-\bar{V})^{2}$ $\rangle$.} 
\label{Fig.3}
\end{center}
\end{figure}

The sharp peak of the 1/f noise allows to determine $T_{c}$ with a high
accuracy; below we identify $T_{c}$ with the temperature of the $S_{V}/V^{2}$
maximum. There is a correlation between the magnitude of the noise peak and
the ratio $\rho(T_{c})/\rho(300K)$. For example, for the sample in Fig. 2,
$\rho$ increases by 4 orders of magnitude with cooling from room temperature
down to $80K$; the normalized noise magnitude at the transition also increases
by a factor of $\sim10^{4}$. Since the high-temperature portion of the
$\rho(T)$ \ dependences is approximately the same for all compounds
$La_{5/8-y}Pr_{y}Ca_{3/8}MnO_{3}$ with $y>0.35$ \cite{masatomo}, the noise peak
is more pronounced for materials with a higher value of $\rho(T_{c})$, e. g.
with a lower $T_{c}$.

\begin{figure}[!hbt]
\begin{center}
\includegraphics[height=5in, width=4in]{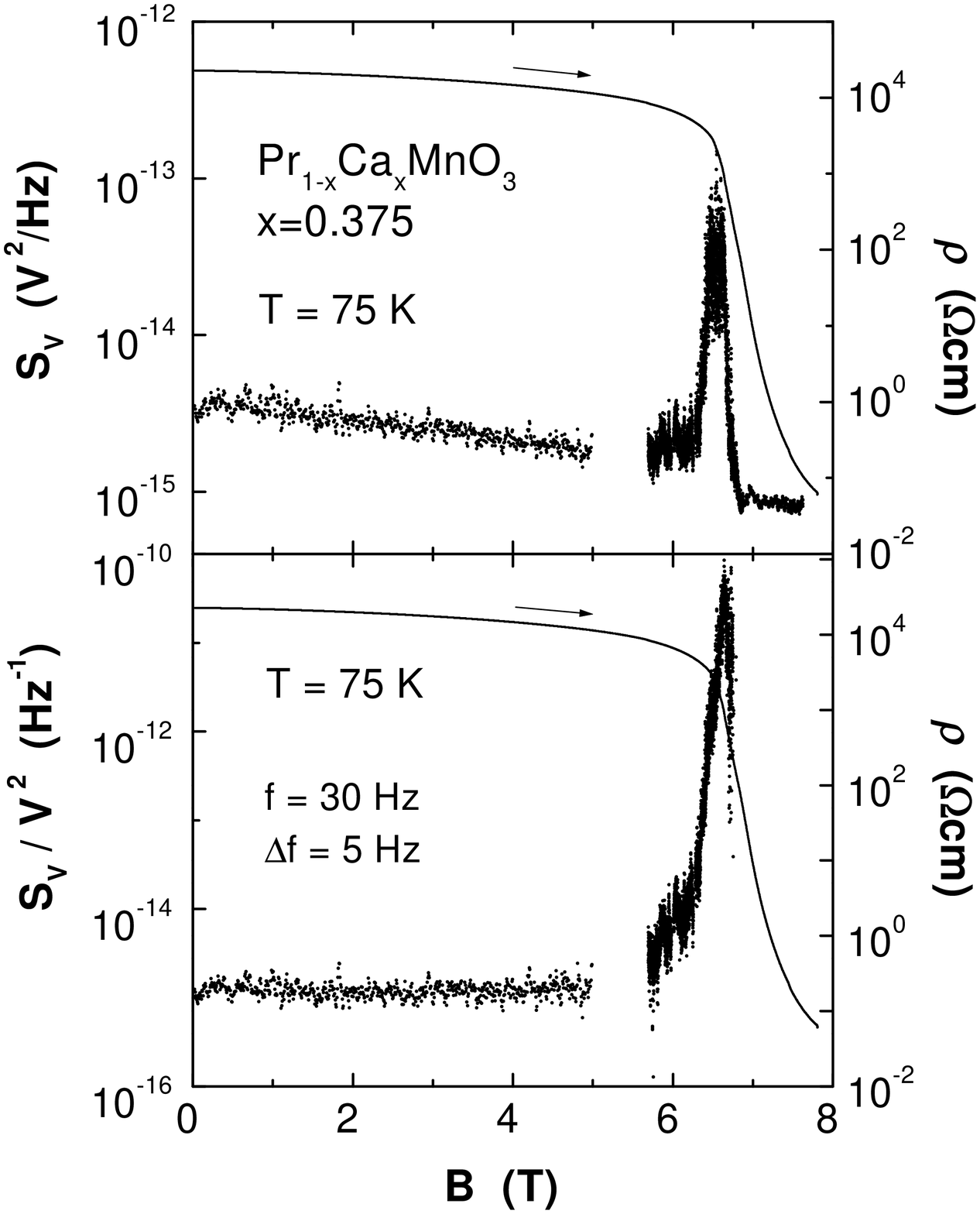}
\caption{Magnetic field dependences of the resistivity (solid lines), the spectral
noise density (dots, the upper panel), and the normalized spectral noise density
corrected on the background (dots, the lower panel) for a polycrystalline sample
$Pr_{1-x}Ca_{x}MnO_{3}$ with $x=0.375$. All the dependences have been measured
at $T=75K$ after zero-field cooling from room temperature. The magnetic field
sweep rate was $2\cdot10^{-4}T/s$.} 
\label{Fig.4}
\end{center}
\end{figure}

Qualitatively similar behavior is observed for the $\Pr_{1-x}Ca_{x}MnO_{3}$
system, where the MIT can be induced by the magnetic field. The upper panel of
Fig. 3 shows the $1/f$ noise versus $T$ for the $\Pr_{0.625}Ca_{0.375}MnO_{3}$
sample at zero magnetic field (where the MIT is absent), and for the field
cooling at $B=6T$ (the MIT is observed at $T=63K$). At $B=0$, the
normalized noise varies weakly with temperature over a range $T=50-80K$.
If the sample is cooled
down at a fixed $B$, the $1/f$ noise increases sharply at the transition.
A ''shoulder'' on the FM side of the $\rho(T)$ dependence
coincides with the second peak of the $1/f$ noise. This shoulder, more or less
pronounced, has been observed for all the samples $\Pr_{1-x}Ca_{x}MnO_{3}$. A
very sharp peak of noise is also observed if the sample is initially
zero-field-cooled down to a certain $T$, and then the MIT is induced by
increasing magnetic field (see Fig. 4). Comparison of Figs. 3 and 4 shows that
the field cooling results in the peak magnitude of the $1/f$ noise
approximately one order of magnitude greater than that for the isothermic
field-induced MIT if the sample was initially zero-field-cooled. In both
cases, the magnitude of the noise peak correlates with sharpness of the
transition. The noise peak is much smaller for compounds with $x<0.35$ and
$x>0.4$, where the resistance drop at the transition is less
dramatic.

Our 1/f noise measurements provide strong evidence of the percolation nature
of the CO-FM transition in the polycrystalline bulk samples of low-$T_{c}$
manganites. Indeed, a diverging behavior of the 1/f noise is typical for the
percolation metal-insulator transition \cite{rammal1,tremblay}. On the insulating
side of the MIT, the $1/f$ noise is relatively small (it is still greater than 
in macroscopically homogeneous metals and semiconductors), and almost temperature-
independent. The concentration of disconnected FM domains increases with approaching
$T_{c}$ on cooling. Switching of these domains between FM and CO states \cite{weis},
results in a rapid growth of the $1/f$ noise with $T\to T_{c}$. This can be especially 
clearly seen for single crystals \cite{podzorov}. The noise intensity
exhibits a giant maximum at $T_{c}$, where the concentration of the FM phase reaches
the percolation threshold, and a cluster of connected FM domains extends through 
the whole sample. Further strengthening of the backbone of the infinite cluster on 
the "metallic" side of the MIT causes a rapid fall of the $1/f$ noise.
Note that $T_{c}$ is lower than the temperature of the maximum resistivity; instead, 
$T_{c}$ coincides with the maximum of $d\rho(T)/dT$. This is expected for a
percolating mixture of two phases where the ''insulating'' phase has a finite
$\rho$ which increases rapidly with cooling.

\begin{figure}[!htb]
\begin{center}
\includegraphics[height=3.5in, width=3.5in]{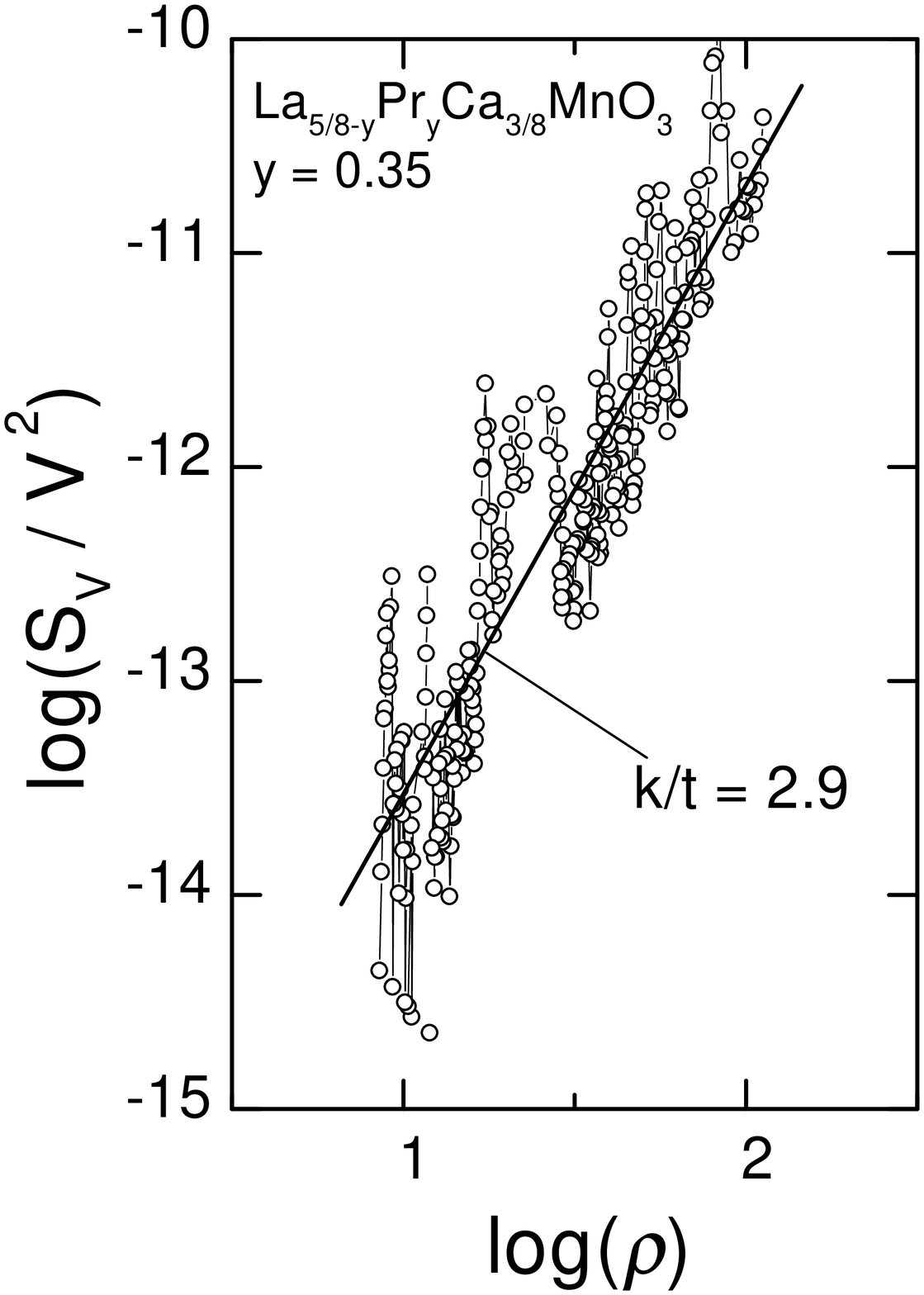}
\caption{The scaling dependence of the normalized spectral density of the 1/f noise 
on $\rho$ for the polycrystalline sample $La_{5/8-y}Pr_{y}Ca_{3/8}MnO_{3}$ with
$y=0.35$ ($f=10Hz$, $\Delta f=1Hz$). Solid line is a power-law fit with the exponent $k/t=2.9$.} 
\label{Fig.5}
\end{center}
\end{figure}

A clearly diverging behavior of the 1/f noise allows to determine $T_{c}$ with
a high accuracy and to perform the scaling analysis of $S_{V}$ and $\rho$ on
the ''metallic'' side of the MIT. In the vicinity of a
percolation metal-insulator transition, the scaling behavior of $\rho$ and
$S_{V}/V^{2}$ is expected \cite{rammal1,kogan,tremblay}:

\begin{equation}
S_{V}/V^{2}\varpropto(p-p_{c})^{-k}\text{ ,}%
\end{equation}

\begin{equation}
\rho\varpropto(p-p_{c})^{-t}\text{ .}%
\end{equation}

Here $p$ is the concentration of the metallic phase, $p_{c}$ is the critical
concentration, $k$ and $t$ are the critical exponents of the noise and 
resistivity, correspondingly. It is convenient to represent $S_{V}/V^{2}$ as a function of
$\rho$ (in this case, no assumption on the value of $p_{c}$ is necessary):

\begin{equation}
S_{V}/V^{2}\varpropto\rho^{k/t}\text{ .}%
\end{equation}

The normalized magnitude of the 1/f noise versus $\rho$ for the
polycrystalline sample $La_{5/8-y}Pr_{y}Ca_{3/8}MnO_{3}$ with $y=0.35$ is shown in the
double-log scale in Fig. 5. Within the experimental accuracy, this
dependence can be fitted by the power law (3) with $k/t=2.9\pm0.5$. For the
other samples, the values of $k/t$ fall into the range $1.2-3$. These values
of $k/t$ are consistent with the result $k/t=2.4$ obtained for the
continuum percolation model of conducting regions, randomly placed in an
insulating matrix (the so-called inverted random-void model)\cite{tremblay}.
Previously, similar values of $k/t=2-3$ have been observed experimentally for
the mixed powders of conducting and insulating materials \cite{rudman,chen}.

For high-quality single crystals of $La_{5/8-y}Pr_{y}Ca_{3/8}MnO_{3}$
($y\sim0.35$) \cite{comment1}, a dramatic increase of
the $1/f$ noise at the transition has been observed as well \cite{podzorov}. 
In contrast to the polycrystalline samples, the temperature
dependence of $\rho$ for single crystals exhibits reproducible steps and switching 
in the vicinity of the MIT. The step-like behavior and switching of $\rho$ at the
transition indicates that the scale of the phase inhomogenuity is considerably greater 
in single crystals than in polycrystalline samples and the contribution of each individual
metalic domain to the transport is noticeble. The percolation approach is not
applicable in this case, because the inhomogeneous system is probed at the scale
comparable with the scale of the phase inhomogenuity. Consequently, the critical
behaviour of the 1/f noise and resistivity could not be observed.

\vspace{0.25in} \noindent CONCLUSIONS \vspace{0.25in}

To summarize, we observed the dramatic peak of the $1/f$ noise in low-$T_{c}$
CMR manganites $La_{5/8-y}Pr_{y}Ca_{3/8}MnO_{3}$ and $Pr_{1-x}Ca_{x}MnO_{3}$
at the transition between the charge-ordered insulating and ferromagnetic
metallic states. The peak value of the normalized noise density is by several
orders of magnitude greater than that in macroscopically homogeneous 
metals and semiconductors.
The combination of these data with the temperature dependence of $\rho$ and
$M$ provides a strong evidence that the metal-insulator transition in these
compounds is, in fact, a percolation crossover in the system of intermixed 
FM-metalic and CO-insulating regions. The metal-insulator transition 
corresponds to the formation of a critical cluster of connected 
metallic FM domains which extends through the whole sample. In other words, 
the metal-insulator transition temperature, or 
the ferromagnetic Curie temperatute $T_{c}$, is the temperature of the percolation 
threshold. The percolation theory describes adequately
the $1/f$ noise in the vicinity of the MIT in polycrystalline samples, where
the size of the FM domains is much smaller than the sample's dimensions. A
well-pronounced step-like temperature dependence of the resistivity, observed
for high-quality single crystals, suggests that the scale of the phase
separation in single crystals is much greater than in polycrystalline samples.

\bigskip\ 

\noindent ACKNOWLEDGEMENTS \vspace{0.25in}

We thank Sh. Kogan, M. Weismann, and V. Kiryukhin for helpful discussions.
This work was supported in part by the NSF grant No. DMR-9802513.


\begin{thebibliography}{99}
\bibitem{comment1}The $Pr$ concentration for the single crystal,
estimated from the $T_{c}(y)$ dependence, was close to 0.35 (the nominal
concentration was $y=0.42$).
\end{thebibliography}
\end{document}